\documentstyle[preprint,aps]{revtex}
\begin{document}
\newcommand{\that}{\hat{\bf t}}
\newcommand{\nhat}{\hat{\bf n}}
\newcommand{\r}{{\bf r}}

\title{A Nonlocal Contour Dynamics Model \\ for Chemical Front Motion}
\author{Dean M. Petrich\footnote{dpetrich@puhep1.princeton.edu}
and Raymond E. Goldstein\footnote{gold@puhep1.princeton.edu}}
\address{Department of Physics, Joseph Henry Laboratories \\
Princeton University, Princeton, NJ 08544}
\maketitle
\begin{abstract}
Pattern formation exhibited by a
two-dimensional reaction-diffusion system
in the fast inhibitor limit is considered from the point of view of
interface motion.  A dissipative nonlocal equation of motion for the
boundary between high and low concentrations of the slow species is
derived heuristically.
Under these dynamics,
a compact domain of high concentration may develop into a space-filling
labyrinthine structure in which nearby fronts
repel.  Similar patterns have been observed
recently by Lee, McCormick, Ouyang, and Swinney in a reacting chemical
system.
\end{abstract}

\newpage
In the study of chemical systems with both reactions and diffusion, one may
discern two broad classes of spatial patterns: {\it extended} and {\it
compact}.
Extended patterns typically arise from supercritical symmetry-breaking
bifurcations \cite{Turing}.  In the two dimensional case, which we
consider here, they
are often regular, periodic structures such as arrays of stripes, discs, or
hexagons \cite{Swinney}.
Compact patterns,
or {\it localized states}, appear in systems with subcritical
bifurcations via a nucleation process \cite{Koga}, and typically take
the form of a
single one of the repeating units found in extended systems.
Both classes of patterns are often described in terms of a competition
between two chemical species: an autocatalytic ``activator" and
its ``inhibitor."

Localized states can exhibit a fingering instability
\cite{Ohta}.  One can imagine that these fingers
may grow and branch until a complicated labyrinthine
pattern fills the entire plane.
Such structures may actually be metastable states;
barriers to domain fission may prevent the pattern
from evolving into the ground state (presumably a regular
array of stripes) from an initial condition
which is topologically different.
Qualitatively similar kinds of pattern formation
appear in other systems \cite{Seul_films_Rosensweig_ferropap}.

Complicated, labyrinthine pattern evolution in a chemical system has
recently been observed by Lee, McCormick, Ouyang, and Swinney \cite{Lee} in an
iodate-ferrocyanide-sulfite reaction (see Fig. \ref{fig1}).
The patterns are composed of regions with one of two
different chemical compositions.
This system appears to be bistable; if it is prepared with one of the
two possible uniform compositions, it will persist in that state.
Nontrivial pattern formation requires a
nucleation site in an otherwise uniform background.
These experiments also indicate that
in the formation of patterns, nearby boundaries
or fronts {\it repel} each other.  We will use these
results as a guide to the general
features one would like to see
in a model system \cite{Pearson}.

It is natural to seek a representation of the dynamics of these
and similar patterns
in terms of the interface between domains of different composition; in
order to capture the properties of repulsion and nonintersection, such a
contour dynamics must be nonlocal, coupling segments of the interface
that are distant in arclength yet close in space.
The purpose of this letter is twofold:
to give an intuitive
construction of a fully nonlocal curve
dynamics from a set of
reaction-diffusion equations and to show that
it is useful for describing pattern formation.
The evolution equation is written in terms of the intrinsic geometry of
the curve, and is valid for pattern evolution far
beyond the linear instability of localized states.
In this simplified form, it is easier to identify the
physics responsible for the destabilization of a compact pattern, the tendency
of a pattern to grow or shrink, and
the interaction between different portions
of the interface.
The curve dynamics is studied numerically and is
shown to reproduce the qualitative features of the experimental patterns.

The reaction-diffusion pair studied here is similar to
models of spiral wave formation \cite{spiral_waves}
and nerve impulse propagation \cite{FitzHugh}.
With $u$ the activator and $v$ the inhibitor, we consider
\begin{mathletters}
\label{uvequations}
\begin{equation}
u_t = D \nabla^2 u -{\rm F}'(u) - \rho v~,
\label{uva}
\end{equation}
\begin{equation}
\epsilon v_t = \nabla^2 v - \alpha v + \beta u~.
\label{uvb}
\end{equation}
\end{mathletters}
The nonlinear function ${\rm F}(u)$,with derivative ${\rm F}'$, embodies
the autocatalytic nature of the
activator.  It is typically a polynomial in $u$ with a double well structure
whose minima, not necessarily of equal depth, we label $u_{\pm}$.
Patterns like that in Fig. \ref{fig1} can then form
in which regions of $u \simeq u_{+}$ (e.g. white) are surrounded by a region of
$u
\simeq u_{-}$ (black).
Inhibition of $u$ is achieved for $\rho >0$,
while the couplings $\alpha$ and $\beta$ reflect the self-limiting behavior
of the inhibitor and its stimulation by the activator, respectively.
The small
number $\epsilon$ defines the fast-inhibitor limit.
This limit is {\it opposite} to the limit
assumed in phase-field models \cite{Fife}
and spiral wave dynamics \cite{spiral_waves}.
The fast inhibitor assumption appears to be the simplest assumption that allows
the elimination of one of the fields, giving rise to
spatial nonlocality for the remaining field without introducing temporal
nonlocality.  A similar calculation can be done for
the slow-inhibitor
limit \cite{future}, but due to temporal nonlocality,
the resulting equations
are more complicated,
obscuring the essential physics.

We begin by discussing the interaction of fronts, illustrating
the property of self-avoidance present in the reaction-diffusion
pair (1), for certain parameter ranges.  Figure 2 shows a simulation
of a one-dimensional version of (1) with periodic
boundary conditions.  The patterns are defined by
the sharp interfaces of the $u$ field.  The parameters are such that
the $u \simeq u_{+}$ region expands into the $u \simeq u_{-}$
region.
When the interfaces get too close, the exponential tails of the $v$
field begin to overlap,
causing a repulsion which stabilizes the
final field configuration.
In the derivation
of the two-dimensional contour dynamics, we shall analytically see the
source
of this repulsion.

To derive an interface evolution equation for two-dimensional patterns,
we take the fast-inhibitor limit, setting $\epsilon = 0$ and
thus slaving $v$ to $u$,
\begin{equation}
v({\bf x},t) = \beta
\int\! d^2x' {\cal G}(\vert {\bf x}-{\bf x'}\vert) u({\bf x'})
\end{equation}
where the Green's function ${\cal G}(r) = {\rm K}_0 (r/\xi)/(2\pi)$,
and $\xi=\alpha^{-1/2}$.
Substituting for $v$ in equations (\ref{uva}), we obtain the variational
dynamics
$u_t = -\delta {\cal F}[u]/\delta u ~$,
with
\begin{eqnarray}
{\cal F}[u]&&=\int\! d^2x
\left\{\frac{1}{2}D\left(\bbox{\nabla} u\right)^2 + F(u)\right\}\cr
&&\quad+{\rho \beta\over 2} \int\! d^2x \! \int\! d^2 x'
u({\bf x}) {\cal G}(\vert {\bf x}-{\bf x'}\vert) u({\bf x'}).
\label{uLyapunov}
\end{eqnarray}
The functional ${\cal F}$ decreases monotonically in time,
possibly reaching a local minimum.
For $\epsilon >0$, there is no such ${\cal F}$.
The necessity of a large-amplitude perturbation for the initiation of
nontrivial pattern formation follows directly from the bistability
of $F(u)$ and the variational form of the dynamics.

We would like to find an equation of motion for the interface
${\bf r}(s)$ in the form $\r_t=-
\delta {\cal F}[{\bf r}]/ \delta {\bf r}$, starting from the equation
$u_t = -\delta {\cal F}[u]/\delta u ~$.
To determine the boundary functional ${\cal F}[{\bf r}]$ from ${\cal F}[u]$
in (\ref{uLyapunov}),
we restrict ourselves to studying the dynamics of a single island of
$u_{in} \simeq u_{+}$ in an infinite sea of $u_{out} \simeq u_{-}$.
We make two assumptions to evaluate the term involving $\bbox{\nabla}u$:
the profile is sharp, so that the gradients
are localized on the boundary $\Gamma$, and the curvature of the boundary
does not significantly affect the interface profile.
The derivation of the
contour dynamics using
asymptotic expansions will be discussed elsewhere \cite{future}.
We proceed here
with a heuristic derivation because it yields greater physical
insight.
The integral of $\left(\bbox{\nabla} u\right)^2$
is then proportional to the perimeter of the shape, the
constant of proportionality being approximately $D(u_{in}-u_{out})^2/l$,
with $l$ a characteristic
length scale of the profile.  One cannot determine $l$ further
without specifying the precise form of $F$.

To evaluate the remaining terms in ${\cal F}$, we take
$u$ to be piecewise constant.  The first term gives
$\int\! F(u) = F(u_{in}) \int_{in} ~+ ~~F(u_{out})\int_{out} ~$,
where $in$ and $out$ refer to the areas inside and outside of $\Gamma$.
The term involving ${\cal G}$ can be written as
\begin{eqnarray}
\int\!\!\int\! u {\cal G} u
&&=u_{in}^2 \int_{in}\! \int_{in} \! {\cal G} ~+~~u_{out}^2 \int_{out}
\! \int_{out} \! {\cal G} \cr
&&\quad +2u_{in} u_{out} \int_{in}\! \int_{out}\! {\cal G}~.
\label{details}
\end{eqnarray}
This may be further simplified by an application of Stokes' theorem
and the
defining relation for ${\cal G}$,
$\left(\nabla^2-\xi^{-2}\right){\cal G}({\bf x})=-\delta({\bf x})$.
Apart from an unimportant
constant, the energy is then
\begin{equation}
{\cal F}[{\bf r}]=\Pi A + \gamma L
-\frac{\rho \beta \xi^2 \Delta^2}{2}
\oint\! ds\oint\! ds' \that \cdot \that'
{\cal G}\left(R\right),
\label{energyfunc}
\end{equation}
where $\gamma \simeq D \Delta^2/2l$,
\begin{equation}
\Pi \simeq F(u_{in})-F(u_{out}) +
\left(\rho\beta \xi^2/2\right)\left(u_{in}^2-u_{out}^2 \right)~,
\label{coefficients}
\end{equation}
and $\Delta=u_{in}-u_{out}$, ${\bf R}=\r(s)-\r(s')$, $R=\vert{\bf R}\vert$.
We interpret $\gamma$ as a line tension
(associated with the boundary length $L$) and $\Pi$ as
a pressure (associated with the enclosed area $A$).
The pressure, which can be of either sign, reflects the difference
between the values of the energy ${\rm F}(u)$ inside and outside $\Gamma$.
Note from the definition of $\Pi$ that the coupling of $u$ to $v$
has created an effective
energy $F(u) + (\rho \beta\xi^2/2 )u^2$ whose minima determine the values
$u_{in}$ and $u_{out}$.
The nonlocal terms in (\ref{energyfunc}) (and
(\ref{normal_velocity}), below)
have the appearance of a self-induction interaction, although
${\cal G}$ provides an exponential screening.
While at first sight somewhat unusual, a coupling between tangent vectors
appears in other systems with
piecewise constant fields \cite{Dritschel_pra,REGDMP}.

Next we determine the equation of motion of ${\bf r}$ from that of $u$.
This has two parts: (i) relating the time derivatives
$u_t$ and $\r_t$, and (ii)
relating the functional derivatives $\delta/\delta u$ and
$\delta/\delta \r$.
Since $u_t$ is nonzero only
near the boundary, the approximation
$u_t \simeq (\Delta/l)\nhat \cdot {\bf r}_t$
is valid.
Likewise, the variation $\delta / \delta u$ is large only near the boundary,
suggesting the identification \cite{REGDMP}
$(\delta/ {\delta u}) \to
(l/ \Delta) \nhat \cdot (\delta/\delta {\bf r})~$.
Rescaling time by
$\left({\Delta/l}\right)^{-2}$, the equation of motion is
\begin{eqnarray}
\nhat\cdot \r_t &=& -\Pi -\gamma\kappa(s)  \cr
&&~ -
\rho \beta \xi^2 \Delta^2
\oint\! ds'{\bf {\hat R}} (s,s')
\times \that(s') \, \,{\cal G}'(R).
\label{normal_velocity}
\end{eqnarray}
The prime on ${\cal G}$ indicates a derivative with respect to
$R$, and
the cross product is a scalar in two dimensions.

Note that the repulsion between adjacent fronts depends only on the
fact that ${\cal G} > 0,$ and {\it not} on the specific form of
${\cal G}$ (as can be seen from (\ref{energyfunc}) and the
variational form of the equation of motion).
This is important, because in a more exact derivation
of the contour dynamics that takes the interface profile
into account, the function appearing in
equation (\ref{normal_velocity}) may not be ${\cal G}$, but
rather one with a less singular behavior
as $R \rightarrow 0$ \cite{ref_remark}.
Note however that ${\cal G}$ has only
a logarithmic, integrable singularity at the origin, so the
dynamics derived here is well
defined.  Even in the presence of such a
cutoff in ${\cal G}$ the
dynamics would not be qualitatively affected, since the
sign of ${\cal G}$ alone determines whether fronts are attracted
or repelled and the large $R$ behavior of the purported
function would be unchanged
(i.e., exponential decay).

By further redefining the time scale such that the coefficient of
the nonlocal term in (\ref{normal_velocity}) is unity,
one finds only
three relevant parameters remain: $\xi$ and
a rescaled $\gamma$ and $\Pi$.
The model (\ref{normal_velocity}) was investiged numerically with
a pseudospectral technique.  Figure \ref{fig2} shows the
evolution of a circle seeded with
sinusoidal perturbations.  A complicated labyrinth
forms due to the repeated fingering of the boundary.
The interface never crosses itself, due to the repulsion
between the adjacent fingers. The competition between the inward pressure
$\Pi$ and the repulsion within a finger sets the finger width, while
the interfinger distance is set by the inhibitor length scale $\xi$ alone.
The nonlocal nature of the dynamics
severely limits the time scale over which the evolution may be followed
numerically; with further computation, the pattern in Fig. \ref{fig2} would
continue to evolve beyond the final picture.

Figure \ref{fig3} shows the evolution of an island with a larger value of $\Pi$
than that in Fig. \ref{fig2},
implying a larger energy
difference between the two possible homogeneous states of $u$.
The area enclosed by the interface changes dramatically over time,
the pattern simply shrinking to a circle.
This shrinkage of a fingered structure has also been seen in the work
of Lee, {\it et al.} \cite{Lee}.

To gain insight into the basic mechanism of the fingering instability,
we develop an approximate local dynamics valid
when $\xi \kappa \ll 1$ and the shape is
approximately circular \cite{Duplantier}.
Observe that the most important contribution to the double integral in
(\ref{energyfunc}) is from the
region $\vert s'-s\vert \le \xi$.
Near the point $\r(s)$, expand the scalar product as
$\that(s)\cdot \that(s')\simeq 1-(1/2)(s-s')^2\kappa(s)^2+\cdots~$, then
perform the $s'$ integral over the extended region $[-\infty,\infty]$
to obtain an effective local
energy functional ${\cal F}_{v}$:
\begin{equation}
{\cal F}_{v}\simeq \oint\! ds\left\{\gamma_{v}+{1\over 2}k_c(\xi\kappa)^2+
{\cal O}\left((\xi\kappa)^4\right)\right\}.
\end{equation}
The first term contributes to an effective line tension, while the
second we recognize from elasticity theory as the bending energy
of a rigid rod.
A detailed calculation yields
$\gamma_{v}= - \rho \xi^3 /4 $ and $k_c= \rho \xi^3 / 8$.
While $k_c$ is positive, $\gamma_{v}$ is {\it negative}.
The effective line tension $\tilde \gamma =
\gamma + \gamma_v$ can be negative, favoring proliferation of the interface.
Again, note that the sign of $\gamma_v$ is opposite that of ${\cal G}$,
and hence the physics of fingering depends only on the sign of ${\cal G}$
and not its specific functional form.

Under the local approximation, the interface motion is
\begin{equation}
\nhat\cdot \r_t\simeq -\Pi-\tilde\gamma \kappa
-k_c\xi^2\left(\kappa_{ss}+{1\over 2}\kappa^3\right)~,
\end{equation}
similar to ``geometric"
models \cite{Brower} of
crystal growth.
In a linear stability analysis
of a circular shape, the growth rate $\sigma_n$ of the
$n^{\rm th}$ mode is $\sigma_n \sim
- \tilde \gamma n^2 - k_c\xi^2 n^4$. For $\tilde \gamma < 0$, there is
a band of unstable modes whose maximum
extent is limited by the rigidity.
In the numerical studies shown here, the
effective tension is negative in both cases.  In Fig. \ref{fig2}, the
inward pressure is insufficient to prevent the proliferation of the
interface, while in Fig. \ref{fig3} the larger pressure overcomes the
negative surface tension.

In summary, starting from a reaction-diffusion system in
two dimensions, we have constructed a boundary dynamics useful in
describing pattern formation.
Possible extensions include studying inertial effects from
higher order terms in $\epsilon$, and the application of these methods to
higher-dimensional problems.
It is also of interest to determine if the particular chemical kinetics
in the experiments of Lee, {\it et al.}
may indeed be mapped onto the interface model developed here.
Finally, it is possible to extend these methods to study the
more physically relevant slow-inhibitor problem \cite{future}.
Perhaps issues such as spiral wave
stability \cite{spiral_waves,Barkley} could be given
a more intuitive interpretation when studied from a geometric viewpoint.

We are grateful to K.J. Lee, W.D. McCormick, Q. Ouyang, and H.L. Swinney
for communication of their results prior to publication and for kindly
providing Fig. \ref{fig1}, and thank V. Hakim and
S. Leibler for important discussions on localized states, and
D. Barkley, S. Erramilli, M.J. Shelley, and A.T. Winfree for discussions and
correspondence.
This material is based upon work supported under an N.S.F.
Graduate Fellowship (DMP), N.S.F. Presidential Faculty Fellowship
Grant DMR-9350227 and the Alfred P. Sloan
Foundation (REG).

\begin{figure}
\caption{Pattern formation in the chemical system of Lee, McCormick,
Ouyang, and Swinney \protect{\cite{Lee}.  White and black regions correspond,
respectively, to low and high pH, made visible with a pH indicator.
Figure courtesy of Lee, {\it et al.} [6].}
\label{fig1}}
\end{figure}

\begin{figure}
\caption{Space-time portrait of the interaction of two fronts, from
numerical solution of Eq. 1 with $\epsilon=0.008$,$D=0.2$, $\rho=0.1$,
$\alpha=\beta=0.2$, $F'(u)=au+bu^2+du^3$, with $a=0.23$, $b= -1.23$, and
$c=1.0$.
Solid lines
show $u(x,t)$, dashed lines are $v(x,t)$, with time increasing
upward.\label{fig1a}}
\end{figure}

\begin{figure}
\caption{Numerical solution of (\protect{\ref{normal_velocity}}) with
$\gamma=2.0$, $\Pi=0.02$, and $\xi=1.0$.  The initial condition is a
perturbed circle of radius $15$, and the time interval between shapes
is $12$. \label{fig2}}
\end{figure}

\begin{figure}
\caption{Shrinkage of a domain with $\gamma=2.0$, $\Pi=2.0$, and $\xi=1.0$.
The time interval between shapes is $1$.
\label{fig3}}
\end{figure}

\end{document}